\documentclass[letterpaper, 10 pt, conference]{ieeeconf} 

\IEEEoverridecommandlockouts
\usepackage{cite}
\usepackage{epsf}
\usepackage{psfrag}
\usepackage{amsmath,amssymb,amsfonts}
\usepackage{algorithmic}
\usepackage{graphicx}
\usepackage{textcomp}
\usepackage{xcolor}
\newcommand{\E}{\mathrm{E}}

\newcommand{\Var}{\mathrm{Var}}

\def\BibTeX{{\rm B\kern-.05em{\sc i\kern-.025em b}\kern-.08em
		T\kern-.1667em\lower.7ex\hbox{E}\kern-.125emX}}
\newtheorem{theorem}{Theorem}

\newtheorem{lemma}{Lemma}
\newtheorem{definition}{Definition}

\begin{document}
	
\title{\LARGE \bf
	Empirical Coordination with Multiple Descriptions
}

\author{Michail Mylonakis$^{1}$, Photios A. Stavrou$^{1}$ and Mikael Skoglund$^{1}$
	\thanks{$^{1}$ The authors are with the Division of Information Science and Engineering,
		KTH Royal Institute of Technology,
		{\{mmyl, fstavrou, skoglund\}@kth.se}.%
	%
}}
	
	
	\maketitle
\begin{abstract}
	 We extend the framework of empirical coordination to a distributed setup where for a given action by nature, multiple descriptions of the action of the decoder are available. We adopt the coding strategy applied by El Gamal and Cover in \cite{gamal:1982} to get a lower bound of the coordination region.  Then, we improve this region by applying the coding scheme applied by Zhang and Berger in \cite{zhang:1987}.
\end{abstract}

\section{Introduction}

The development of information technology and the need
for wireless connectivity in networks have enabled a plethora of new transformative technologies and applications. On the level of information theory, this development necessitates the need for distributed coding between the nodes of a network depending on the action taken whereas at the same time, it is desirable for some applications, such as, parallel computing across large scale networks, a network of computers to be interconnected under coherent strategies while distributing the work load across the participating machines (i.e., machine-type communication).

In this paper, we apply ideas from a distributed source coding setup called the {\it multiple description problem} (MDP) \cite{wolf:1980} to the framework of {\it empirical coordination using a fidelity criterion} that was recently introduced in \cite{mylonakis:2019}.

The idea of MDP was formalized in \cite{wolf:1980} and it is concerned with lossy encoding of information for transmission over an unreliable (and possibly digital) multi-channel communication system. The receiver knows which subset of the channels is working whereas the transmitter does not. The problem boils down to the design of an MD system which, for given channel rates, minimizes the distortions due to reconstruction of the source using information from any subsets of the channels. The first achievability rate-region for MDP using two channels was established in \cite{gamal:1982} and further generalized in \cite{zhang:1987}. A detailed and simple exposition of these results is given in \cite{moser:2019}. The rate region in \cite{gamal:1982} was shown to be optimal for memoryless Gaussian sources and mean squared error distortion in \cite{ozarow:1980}. This result was later extended to create high-rate bounds for stationary and smooth sources in \cite{zamir:1999} and for stationary Gaussian processes in \cite{dragotti:2002}. Bounds to the optimal rate distortion region for MDP for more than two channels are studied in \cite{venkataramani:2003,pradhan:2004,puri:2005,viswanatha:2016}.

 \begin{figure}
	\begin{center}
		
		\includegraphics[width=4cm,height=3cm,keepaspectratio]{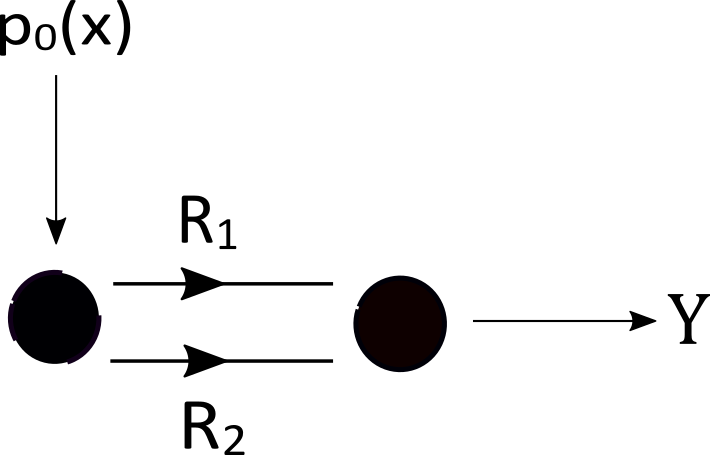}
	\end{center}
	\caption{System model.}
	\label{fig1}
\end{figure}

The notion of empirical coordination was introduced in \cite{cuff:2010} and it is achieved if the joint type, measured by total variation distance, of the actions (or nodes) in a network is close to the desired distribution, in probability. This kind of coordination has been studied in various settings, see e.g., \cite{bereyhi:2013} and merged with ideas from other research fields, such as game theory \cite{letreust:2016} and networked control systems \cite{sharieepoorfard:2018}. The framework of empirical coordination of \cite{cuff:2010} was recently extended to the more general framework of empirical coordination subject to fidelity also termed ``imperfect empirical coordination'' by \cite{mylonakis:2019} who was inspired by the work of \cite{kramer:2007}. Following \cite{mylonakis:2019}, imperfect empirical coordination is established if the total variation between the joint type of the actions in a network comes close, on average, to a desired distribution within distance pre-specified by a threshold $\Delta$. Clearly, if $\Delta=0$, then, we obtain the empirical coordination in the sense of \cite{cuff:2010}.

In this work, we extend the idea of MDs to the framework of imperfect empirical coordination, using the model of Fig. \ref{fig1}. In the system model of Fig. \ref{fig1}, we are given the action of a random variable $X$ (possibly by nature) that induces the distribution $p_0\left(x\right)$. Moreover, the action of $Y$ is produced using two available rate-limited noiseless communication links. 
We seek, the amount of the communication that is required such that the average distance between the joint type of the actions and the desired joint distribution $p_0\left(x\right)p_{Y|X}\left(y|x\right)$ to be smaller than a certain thresold $\Delta_i$, for $i=1,2,12$, depending on whether the first, the second, or both indices arrive at the decoder, respectively. The system in Fig. \ref{fig1} can be used in practice. For instance, consider a network where due to packet loss some part of the message does not arrive at the receiver. A traditional coordinated system will fail, while a multiple
description system can still coordinate the nodes, with less
accuracy. 
	\section{Definitions}
    In this section, we describe formally our problem.
	We begin with some mathematical preliminaries and the definition of the $\Delta$-neighborhood. 
	\begin{definition}[Joint type] The joint type $P_{x^n,y^n}$ of a tuple of sequences $\left(x^n,y^n\right)$ is the empirical probability mass function, given by
		\begin{equation*}P_{x^n,y^n}\left(x,y\right)\triangleq \frac{1}{n}\sum_{i=1}^n{\mathbf 1\big(\left(x_i,y_i\right)=\left(x,y\right)\big)},\end{equation*}
		for all $\left(x,y\right)\in \mathbb{X}\times \mathbb{Y}$, where $\mathbf 1$ is the indicator function.
	\end{definition}
	
	\begin{definition}[Total variation] The total variation between two probability mass functions (PMF) is given by \begin{equation*}\|p\left(x,y\right)-q\left(x,y\right)\|_{TV}\triangleq\frac{1}{2}\sum_{x,y}{|p\left(x,y\right)-q\left(x,y\right)|}.\end{equation*}
	\end{definition}
	\begin{definition}[$\Delta$-neighborhood]
		The $\Delta$-neighborhood of a PMF $p\left(x,y\right)$ is defined as
		\begin{IEEEeqnarray*}{rCl} N_{\Delta}\big(p\left(x,y\right)\big)\}\triangleq \big\{q(x,y):\|p\left(x,y\right)-q\left(x,y\right)\|_{TV}\leq\Delta\big\}.
		\end{IEEEeqnarray*}
	\end{definition}
	\begin{definition}[Coordination code]
		The $\left(e^{nR_1},e^{nR_2},n\right)$ coordination code for our set-up consists of five functions-two encoding functions
		\begin{equation*}
		i_i:\mathbb X^n
		\rightarrow\left\{1,\dots,e^{nR_i}\right\},\quad i=1,2,
		\end{equation*} 
		and three decoding functions 
		\begin{IEEEeqnarray*}{rCl}
		y_i^n&:&\left\{1,\dots,e^{nR_i}\right\}
		\rightarrow \mathbb Y^n,\quad $i=1,2$,\\
		y_{12}^n&:&\left\{1,\dots,e^{nR_1}\right\}\times \left\{1,\dots,e^{nR_2}\right\}%
		\rightarrow \mathbb Y^n.
		\end{IEEEeqnarray*} 
	\end{definition}
	 \begin{definition}[Achievability] 
		A desired PMF $p_{X,Y}\left(x,y\right)\triangleq p_{0}\left(x\right)p_{Y|X}\left(y|x\right)$ is achievable for $\Delta_1\Delta_2\Delta_{12}$-empirical coordination with the rate-pair $\left(R_1,R_2\right)$ if there is an $N$ such that for all $n>N$, there exists a coordination code $\Big(e^{nR_1},e^{nR_2},n\Big)$ such that
		\begin{IEEEeqnarray*}{rCl}
			\mathbb E\big\{\|P_{x^n,y^n}\left(x,y\right)-p_{0}\left(x\right)p_{Y|X}\left(y|x\right)\|_{TV}\big\}&\leq& \Delta_i,\\
		    \end{IEEEeqnarray*}for $i=1,2,12$, depending on whether the first, the second, or both indices arrive at the decoder, respectively.
\end{definition}
\begin{definition}[Multiple-rate-distortion-coordination region]
The Multiple-rate-distortion-coordination region $R_{p_0}^I$ for every source PMF $p_0\left(x\right)$ and for every conditional PMF $p_{Y|X}\left(y|x\right)$ is defined as:
\begin{multline*}
R_{p_0}^I\big(p_{Y|X}\left(y|x\right)\big)\\ \quad \triangleq
\mathbf{Cl}\left.
\left\{ \,
\begin{IEEEeqnarraybox}[
\IEEEeqnarraystrutmode
\IEEEeqnarraystrutsizeadd{1pt}
{1pt}][c]{l}
\left(R_1,R_2,\Delta_1,\Delta_2,\Delta_{12}\right):\\
p_{0}\left(x\right)p_{Y|X}\left(y|x\right) \text{is achievable}\\\text{for $\Delta_1\Delta_2\Delta_{12}$-empirical coordination}\\\text{at rates $\left(R_1,R_2\right)$}
\end{IEEEeqnarraybox}\right\}.
\right.
\end{multline*}
\end{definition}
\section{Main Results}
In this section, we state the main results of this paper. Their detailed derivations are presented in section \ref{proofs}.
\begin{theorem}
The following region is a subset of the Multiple-rate-distortion-coordination region $R_{p_0}^I\big(p_{Y|X}\left(y|x\right)\big)$ for the source PMF $p_0\left(x\right)$ and for every conditional PMF $p_{Y|X}\left(y|x\right)$:
\begin{multline*} R_{p_0}^I\big(p_{Y|X}\left(y|x\right)\big) \supseteq\\ \bigcup_{\substack{p_{\hat{Y}^{\left(1\right)},\hat{Y}^{\left(2\right)},\hat{Y}^{\left(12\right)}|X}:\\ p_0\left(x\right)p_{\hat{Y}^{\left(1\right)}|X}\left(y|x\right)\in N_{\Delta_1}\left(p_{X,Y}\left(x,y\right)\right),\\
		p_0\left(x\right)p_{\hat{Y}^{\left(2\right)}|X}\left(y|x\right)\in N_{\Delta_2}\left(p_{X,Y}\left(x,y\right)\right),\\
		p_0\left(x\right)p_{\hat{Y}^{\left(12\right)}|X}\left(y|x\right)\in N_{\Delta_{12}}\left(p_{X,Y}\left(x,y\right)\right)}} R\left(p_0p_{\hat{Y}^{\left(1\right)},\hat{Y}^{\left(2\right)},\hat{Y}^{\left(12\right)}|X}\right),
\end{multline*}
where
\begin{multline*}
R\left(p_0 p_{\hat{Y}^{\left(1\right)},\hat{Y}^{\left(2\right)},\hat{Y}^{\left(12\right)}|X}\right)\\\quad \triangleq
\left.
\left\{ \,
\begin{IEEEeqnarraybox}[
\IEEEeqnarraystrutmode
\IEEEeqnarraystrutsizeadd{1pt}
{1pt}][c]{l}
\left(R_1,R_2,\Delta_1,\Delta_2,\Delta_{12}\right):\\	
R_1\geq I\left(X;\hat{Y}^{\left(1\right)}\right),\\
R_2\geq I\left(X;\hat{Y}^{\left(2\right)}\right),\\
R_1+R_2\geq I\left(X;\hat{Y}^{\left(1\right)},\hat{Y}^{\left(2\right)},\hat{Y}^{\left(12\right)}\right)\\+I\left(\hat{Y}^{\left(1\right)};\hat{Y}^{\left(2\right)}\right)
\end{IEEEeqnarraybox}\right\}.
\right.
\end{multline*}
\label{th:theorem1}

\end{theorem}
\begin{theorem}
The following region is a subset of the Multiple-rate-distortion-coordination region $R_{p_0}^I\big(p_{Y|X}\left(y|x\right)\big)$ for the source PMF $p_0\left(x\right)$ and for every conditional PMF $p_{Y|X}\left(y|x\right)$:
\begin{multline*} R_{p_0}^I\big(p_{Y|X}\left(y|x\right)\big)\supseteq\\  \bigcup_{\substack{p_{U,\hat{Y}^{\left(1\right)},\hat{Y}^{\left(2\right)},\hat{Y}^{\left(12\right)}|X}:\\p_0\left(x\right)p_{\hat{Y}^{\left(1\right)}|X}\left(y|x\right)\in N_{\Delta_1}\left(p_{X,Y}\left(x,y\right)\right),\\p_0\left(x\right)p_{\hat{Y}^{\left(2\right)}|X}\left(y|x\right)\in N_{\Delta_2}\left(p_{X,Y}\left(x,y\right)\right),\\p_0\left(x\right)p_{\hat{Y}^{\left(12\right)}|X}\left(y|x\right)\in N_{\Delta_{12}}\left(p_{X,Y}\left(x,y\right)\right)}} R\left(p_0p_{U,\hat{Y}^{\left(1\right)},\hat{Y}^{\left(2\right)},\hat{Y}^{\left(12\right)}|X}\right),
\end{multline*}
where
\begin{multline*}
R\left(p_0 p_{U,\hat{Y}^{\left(1\right)},\hat{Y}^{\left(2\right)},\hat{Y}^{\left(12\right)}|X}\right)\\\quad \triangleq
\left.
\left\{ \,
\begin{IEEEeqnarraybox}[
\IEEEeqnarraystrutmode
\IEEEeqnarraystrutsizeadd{1pt}
{1pt}][c]{l}
\left(R_1,R_2,\Delta_1,\Delta_2,\Delta_{12}\right):\\	
R_1\geq I\left(X;\hat{Y}^{\left(1\right)},U\right),\\
R_2\geq I\left(X;\hat{Y}^{\left(2\right)},U\right),\\
R_1+R_2\geq I\left(X;U\right)\\+I\left(X;\hat{Y}^{\left(1\right)},\hat{Y}^{\left(2\right)},\hat{Y}^{\left(12\right)},U\right)\\+I\left(\hat{Y}^{\left(1\right)};\hat{Y}^{\left(2\right)}|U\right)
\end{IEEEeqnarraybox}\right\}.
\right.
\end{multline*}	
\label{th:theorem2}
\end{theorem}
\section{Proofs}
\label{proofs}
\subsection{Proof of Theorem \ref{th:theorem1}}
First, we give the proof of Theorem \ref{th:theorem1} and, then, we sketch the proof of Theorem \ref{th:theorem2}.
\begin{itemize}
\item{\textit{Setup}:} Choose $\epsilon_1,\epsilon_2>0$, a PMF $p_{\hat{Y}^{\left(1\right)},\hat{Y}^{\left(2\right)},\hat{Y}^{\left(12\right)}|X}$ and compute the marginals $p_{\hat{Y}^{\left(1\right)}},p_{\hat{Y}^{\left(2\right)}},p_{\hat{Y}^{\left(12\right)}}$ as marginal distributions of $p_Xp_{\hat{Y}^{\left(1\right)},\hat{Y}^{\left(2\right)},\hat{Y}^{\left(12\right)}|X}$.	\item{\textit{Codebook design}:} Generate $\lfloor e^{n\left(R_i+\epsilon_i\right)} \rfloor$ length-$n$ codewords $\hat{\mathbf Y}^{\left(i\right)}\left(w_i\right), w_i=1,\dots, \lfloor e^{n\left(R_i+\epsilon_i\right)} \rfloor$, by choosing each of the $n\lfloor e^{n\left(R_i+\epsilon_i\right)} \rfloor$ symbols $\hat{Y}^{\left(i\right)}_k\left(w_i\right)$ independently at random according to $p_{\hat{Y}^{\left(i\right)}}$ for $i=1,2$. For each pair $\left(w_1,w_2\right)$, generate a codeword $\hat{\mathbf Y}^{\left(12\right)}\left(w_1,w_2\right)$ by choosing its $k$th component $\hat{Y}^{\left(12\right)}_k\left(w_{1},w_2\right)$ independently at random according to $p_{\hat{Y}^{\left(12\right)}|\hat{Y}^{\left(1\right)},\hat{Y}^{\left(2\right)}}\left(\cdot\big|\hat{Y}^{\left(1\right)}_k\left(w_{1}\right),\hat{Y}^{\left(2\right)}_k\left(w_2\right)\right)$. 
\item{\textit{Encoder Design}:} For a given sequence $\mathbf{x}$, the encoders try to find a pair $\left(w_1,w_2\right)$ such that 
\begin{multline}
\left(\mathbf x,\hat{\mathbf Y}^{\left(1\right)}\left(w_1\right),\hat{\mathbf Y}^{\left(2\right)}\left(w_2\right),\hat{\mathbf Y}^{\left(12\right)}\left(w_1,w_2\right)\right)\\\in \mathbb{A}_{\epsilon}^{\ast\left(n\right)}\big(p_{X,\hat{Y}^{\left(1\right)},\hat{Y}^{\left(2\right)},\hat{Y}^{\left(12\right)}}\big).\label{jtypicality}
\end{multline}
If they find several possible choices, they pick the first. If they find none, they choose $w_1=w_2=1$. The first encoder $i_1$ puts out $w_1$ and the second encoder $i_2$ puts out $w_2$.
\item{\textit{Decoder Design}:} The decoder $\left(y_1^n,y_2^n,y_{12}^n\right)$ consists of three different decoding functions, depending on whether $w_1$,$w_2$, or both 
are received. It puts out $\hat{\mathbf Y}^{\left(i\right)}\left(w_i\right)$ if $w_i$ is received for i=1,2,12.
\item{\textit{Performance Analysis}:} We define $\epsilon^\prime\triangleq\frac{\epsilon}{2|\mathbb Y|^3}$
 and partition the space into three disjoint cases:
(a) $\mathbf x\notin \mathbb{A}_{\epsilon^\prime}^{\ast\left(n\right)}\big(p_{X}\big)$
in which case we for sure cannot find a pair $\left(w_1,w_2\right)$ such that \eqref{jtypicality} is satisfied
(b) $\mathbf x\in \mathbb{A}_{\epsilon^\prime}^{\ast\left(n\right)}\big(p_{X}\big)$ and for every pair $\left(w_1,w_2\right)$, \eqref{jtypicality} is not satisfied
(c) $\mathbf x\in \mathbb{A}_{\epsilon^\prime}^{\ast\left(n\right)}\big(p_{X}\big)$ and exists a pair $\left(w_1,w_2\right)$ such that \eqref{jtypicality} is satisfied.
 Note that the reduced $\epsilon^\prime$ is only necessary in order to be able to rely on the lower bound in Lemma \ref{lem:TB}. Obviously, if $\mathbf X \in \mathbb{A}_{ \epsilon^\prime}^{\ast\left(n\right)}\left(p_{X}\right)$, then, $\mathbf X \in \mathbb{A}_{\epsilon}^{\ast\left(n\right)}\left(p_{X}\right)$.
 We obtain in each of the three scenaria $i=1,2,12$:
\begin{IEEEeqnarray}{rCl}
&&\mathbb E\big\{\|P_{x^n,y^n}\left(x,y\right)-p_{0}\left(x\right)p\left(y|x\right)\|_{TV}\big\}=\nonumber\\&&\underbrace{\mathbb E\big\{\|P_{x^n,y^n}\left(x,y\right)-p_{0}\left(x\right)p\left(y|x\right)\|_{TV}\big|\text{Case a}\big\}}_{\leq TV_{\max}}\nonumber\\&&\times\Pr\left(\text{Case a}\right)\nonumber\\&&+\underbrace{\mathbb E\big\{\|P_{x^n,y^n}\left(x,y\right)-p_{0}\left(x\right)p\left(y|x\right)\|_{TV}\big|\text{Case b}\big\}}_{\leq TV_{\max}}\nonumber\\&&\times\Pr\left(\text{Case b}\right)\nonumber\\&&+\mathbb E\big\{\|P_{x^n,y^n}\left(x,y\right)-p_{0}\left(x\right)p\left(y|x\right)\|_{TV}\big|\text{Case c}\big\}\nonumber\\&&\times\underbrace{\Pr\left(\text{Case c}\right)}_{\leq 1}\nonumber\\&&\leq TV_{\max}\Pr\left(\text{Case a}\right)+TV_{\max}\Pr\left(\text{Case b}\right)\nonumber\\&&+\mathbb E\big\{\|P_{x^n,y^n}\left(x,y\right)-p_{0}\left(x\right)p\left(y|x\right)\|_{TV}\big| \text{Case c}\big\}. \label{eq:totalE}\IEEEeqnarraynumspace
\end{IEEEeqnarray}
By Lemma \ref{lem:TA} (in Appendix), we can bound the probability of Case (a) as $\Pr\left(\text{Case a}\right)\leq \delta_{t}\left(n,\epsilon^\prime,\mathbb{X}\right)$.
To bound the expected distortion in Case (c), we note that if $\left(\mathbf x,\hat{\mathbf Y}^{\left(1\right)},\hat{\mathbf Y}^{\left(2\right)},\hat{\mathbf Y}^{\left(12\right)}\right)$
is jointly typical, then, each pair $\left(\mathbf x,\hat{\mathbf Y}^{\left(1\right)}\right)$,$\left(\mathbf x,\hat{\mathbf Y}^{\left(2\right)}\right)$,$\left(\mathbf x,\hat{\mathbf Y}^{\left(12\right)}\right)$ is jointly typical so \begin{IEEEeqnarray*}{rCl}&&\|P_{x^n,y^n}\left(x,y\right)-p_{0}\left(x\right)p\left(y|x\right)\|_{TV}\nonumber\\&& \leq\|P_{x^n,y^n}\left(x,y\right)-p_{0}\left(x\right)p_{\hat{Y}^{\left(i\right)}|X}\left(y|x\right)\|_{TV}\nonumber\\&& +\|p_{0}\left(x\right)p_{\hat{Y}^{\left(i\right)}|X}\left(y|x\right)-p_{0}\left(x\right)p\left(y|x\right)\|_{TV}\label{eq:trianglein}\leq\frac{\epsilon}{2}+\Delta_{i},\nonumber
\end{IEEEeqnarray*}
by choosing $p_{\hat{Y}^{\left(i\right)}|X}\left(y|x\right)$ such that $p_{0}\left(x\right)p_{\hat{Y}^{\left(i\right)}|X}\left(y|x\right)\in N_{\Delta_{i}}$ for $i=1,2,12$.

Given some $\mathbf x\in \mathbb{A}_{\epsilon^\prime}^{\ast\left(n\right)}\big(p_X\big)$, define $ F\left(w_1,w_2\right)$ to be the event that $w_1$ and $w_2$ give a good choice of codewords, i.e.,
\begin{IEEEeqnarray*}{rCl} &&F\left(w_1,w_2\right)\triangleq \Bigg\{\left(\hat{\mathbf Y}^{\left(1\right)}\left(w_1\right),\hat{\mathbf Y}^{\left(2\right)}\left(w_2\right),\right.\\&& \left.\hat{\mathbf Y}^{\left(12\right)}\left(w_1,w_2\right)\right)\in \mathbb{A}_{\epsilon}^{\ast\left(n\right)}\big(p_{X,\hat{Y}^{\left(1\right)},\hat{Y}^{\left(2\right)},\hat{Y}^{\left(12\right)}}|\mathbf x\big)\Bigg\}.\end{IEEEeqnarray*}
We can write 
$\Pr\left(\text{Case b}\right)=\bigcap_{w_1,w_2}\Big( F^c\left(w_1,w_2\right)\Big)$\newline$
=\Pr\left[K=0\right]$
with $K\triangleq \sum_{w_1,w_2}{\mathbf 1\Big(F\left(w_1,w_2\right)}\Big).$
Note, that if $K=0$, then $\big|K-\mathbb E\left[K\right]\big|=\big|\mathbb E\left[K\right]\big|=\mathbb E\left[K\right]$\newline$\geq \frac{\mathbb E\left[K\right]}{2}.$
So, using this bound, we get 
$\Pr\left(\text{Case b}\right)$\newline$=\Pr\left[K=0\right] \leq\Pr\left[\big|K-\mathbb E\left[K\right]\big|\geq \frac{\mathbb E\left[K\right]}{2}\right]\leq\frac{\Var\left[K\right]}{ \left(\frac{\mathbb E\left[K\right]}{2}\right)^2}$\newline$=\frac{4\Big(\mathbb E\left[K^2\right]-\big(\mathbb E\left[K\right]\big)^2\Big)}{\big(\mathbb E\left[K\right]\big)^2}$,
where the second inequality follows from the Chebyshev inequality. We write down,
\begin{IEEEeqnarray}{c}
\Pr\left(\text{Case b}\right) \leq \frac{4\Big(\mathbb E\left[K^2\right]-\left(\mathbb E\left[K\right]\right)^2\Big)}{\left(\mathbb E\left[K\right]\right)^2}. \label{eq:tprobability}
\end{IEEEeqnarray}
Therefore, it remains to derive some bounds on $\mathbb E\left[K\right]$ and $\mathbb E\left[K^2\right]$. Firstly, $\mathbb E\left[K\right]$:
\begin{IEEEeqnarray}{rCl}
&&\mathbb E \left[K\right]=\mathbb E\Bigg[\sum_{w_1,w_2}{\mathbf 1\Big( F\left(w_1,w_2\right)\Big)}\Bigg]\label{eq:kapamean}\nonumber\\
&&=\sum_{w_1,w_2}{\mathbb E\Big[\mathbf 1\big( F\left(w_1,w_2\right)\big)\Big]}\label{eq:kapamean1}\nonumber\\
&&=\sum_{w_1,w_2}\Big(1\cdot \Pr\big(F\left(w_1,w_2\right)\big)\nonumber\\&&
+0\cdot \Pr\big(F^c\left(w_1,w_2\right)\big)\Big) \nonumber\label{eq:kapamean2}\\
&&=\sum_{w_1,w_2}{\Big( \Pr\big(F\left(w_1,w_2\right)\big)\Big)}\label{eq:kapamean3}\\
&&=\sum_{w_1,w_2}\sum_{\left(\hat{\mathbf y}^{\left(1\right)},\hat{\mathbf y}^{\left(2\right)},\hat{\mathbf y}^{\left(12\right)}\right)\in\mathbb{A}_{\epsilon}^{\ast\left(n\right)}\left(\cdot|\mathbf x\right) }p_{\hat{Y}^{\left(1\right)}}^n\left(\hat{\mathbf y}^{\left(1\right)}\right)\nonumber\\&& \cdot p_{\hat{Y}^{\left(2\right)}}^n\left(\hat{\mathbf y}^{\left(2\right)}\right) p_{\hat{Y}^{\left(12\right)}|\hat{Y}^{\left(1\right)},\hat{Y}^{\left(2\right)}}^n\left(\hat{\mathbf y}^{\left(12\right)}|\hat{\mathbf y}^{\left(1\right)},\hat{\mathbf y}^{\left(2\right)}\right).\IEEEeqnarraynumspace\label{eq:kapamean4}
\end{IEEEeqnarray}
Here, we introduce the shorthand, $\mathbb{A}_{\epsilon}^{\ast\left(n\right)}\left(\cdot|\mathbf x\right)$ for $\mathbb{A}_{\epsilon}^{\ast\left(n\right)}\left(p_{X,\hat{Y}^{\left(1\right)},\hat{Y}^{\left(2\right)},\hat{Y}^{\left(12\right)}}|\mathbf x\right)$.
By repeating application of Lemma \ref{lem:TA} and Lemma \ref{lem:TB} (in Appendix), we bound the previous as follows:
\begin{IEEEeqnarray}{rCl}&&\mathbb E \left[K\right]>\sum_{w_1,w_2}\sum_{\substack{\left(\hat{\mathbf y}^{\left(1\right)},\hat{\mathbf y}^{\left(2\right)},\hat{\mathbf y}^{\left(12\right)}\right)\\\in\mathbb{A}_{\epsilon}^{\ast\left(n\right)}\left(\cdot|\mathbf x\right)}} e^{-n\Big(H\left(\hat{Y}^{\left(1\right)}\right)+\epsilon_m\Big)}\nonumber\\&& \cdot e^{-n\Big(H\left(\hat{Y}^{\left(2\right)}\right)+\epsilon_m\Big)} e^{-n\Big(H\left(\hat{Y}^{\left(12\right)}\big|\hat{Y}^{\left(1\right)},\hat{Y}^{\left(2\right)}\right)+\epsilon_m\Big)}\label{eq:kapamean5}\\&&=\sum_{w_1,w_2}\bigg|\mathbb{A}_{\epsilon}^{\ast\left(n\right)}\left(p_{X,\hat{Y}^{\left(1\right)},\hat{Y}^{\left(2\right)},\hat{Y}^{\left(12\right)}}|\mathbf x\right)\bigg|\nonumber\\&&\cdot e^{-n\Big(H\left(\hat{Y}^{\left(1\right)}\right)+H\big(\hat{Y}^{\left(2\right)}\big)+H\big(\hat{Y}^{\left(12\right)}\big|\hat{Y}^{\left(1\right)},\hat{Y}^{\left(2\right)}\big)+3\epsilon_m\Big)}\label{eq:kapamean6}\\&&>\sum_{w_1,w_2}\left(1-\delta_{t}\right)e^{n\Big(H\left(\hat{Y}^{\left(1\right)},\hat{Y}^{\left(2\right)},\hat{Y}^{\left(12\right)}|X\right)-\epsilon_m\Big)}\nonumber\\&&\cdot e^{-n\Big(H\left(\hat{Y}^{\left(1\right)}\right)+H\big(\hat{Y}^{\left(2\right)}\big)+H\big(\hat{Y}^{\left(12\right)}\big|\hat{Y}^{\left(1\right)},\hat{Y}^{\left(2\right)}\big)+3\epsilon_m\Big)}\label{eq:kapamean7}\\
	&&=\lfloor e^{n\left(R_1+\epsilon_1\right)}\rfloor	\lfloor e^{n\left(R_2+\epsilon_2\right)}\rfloor\left(1-\delta_t\right)\nonumber\end{IEEEeqnarray}\begin{IEEEeqnarray}{rCl}
	&&\cdot \exp\Bigg(n\bigg(H\left(\hat{Y}^{\left(1\right)},\hat{Y}^{\left(2\right)},\hat{Y}^{\left(12\right)}|X\right)-H\left(\hat{Y}^{\left(1\right)}\right) \nonumber\\&&-H\left(\hat {Y}^{\left(2\right)}|\hat {Y}^{\left(1\right)}\right)-H\big(\hat{Y}^{\left(12\right)}\big|\hat{Y}^{\left(1\right)},\hat{Y}^{\left(2\right)}\big)\nonumber\\&&+H\left(\hat {Y}^{\left(2\right)}|\hat {Y}^{\left(1\right)}\right)-H\left(\hat{Y}^{\left(2\right)}\right)-4\epsilon_m\bigg)\Bigg)\label{eq:kapamean8}\\
&&=\exp\Bigg(n\bigg(R_1+R_2+\epsilon_1+\epsilon_2\nonumber\\&& -I\Big(X;\hat{Y}^{\left(1\right)},\hat{Y}^{\left(2\right)},\hat{Y}^{\left(12\right)}\Big)  -I\Big(\hat{Y}^{\left(1\right)};\hat{Y}^{\left(2\right)}\Big)-\delta_1\bigg)\Bigg).\IEEEeqnarraynumspace\label{eq:kapamean9}
\end{IEEEeqnarray}
Next, we tackle $\mathbb E\left[K^2\right]$:
\begin{IEEEeqnarray}{rCl}&& \mathbb E\left[K^2\right]=\mathbb E\Bigg[\sum_{w_1,w_2}{\mathbf 1\Big(F\left(w_1,w_2\right)\Big)} \sum_{v_1,v_2}{\mathbf 1\Big( F\left(v_1,v_2\right)\Big)}\Bigg]\nonumber\label{eq:skapamean}\\&&=\sum_{w_1,w_2}\sum_{v_1,v_2}\mathbb E\Bigg[\mathbf 1\Big(F\left(w_1,w_2\right)\Big) \mathbf 1\Big(F\left(v_1,v_2\right)\Big)\Bigg]\nonumber\label{eq:skapamean1}\\&&=\sum_{w_1,w_2}\sum_{v_1,v_2}\Pr\Big(F\left(w_1,w_2\right)\cap  F\left(v_1,v_2\right)\Big)\nonumber\label{eq:skapamean2}\\&&=\sum_{\Theta\subseteq\left\{1,2\right\}}\sum_{\substack{w_1,w_2,\\v_1,v_2\\ \text{with overlap}\\ \Theta}}\Pr\Big( F\left(w_1,w_2\right)\cap F\left(v_1,v_2\right)\Big),\IEEEeqnarraynumspace  \label{eq:skapamean3}
\end{IEEEeqnarray}
where in the last step we distinguish four cases of whether $w_i=v_i$ or not. These cases are described by the four possible subsets of $\left\{1,2\right\}$: $\Theta=\left\{1,2\right\},\Theta=\left\{1\right\},\Theta=\left\{2\right\},\Theta=\emptyset$. For $i\in \Theta$, we have $w_i=v_i$, whereas for the remaining indices $i\in \Theta^c$ we have $w_i\neq v_i$. Next, we go through these four cases. 

\textbf{Case $\mathbf{\Theta=\left\{1,2\right\}}$:} We have $w_1=v_1$ and $w_2=v_2$. Using a derivation similar to the one that gave \eqref{eq:kapamean3}-\eqref{eq:kapamean9} we get
\begin{IEEEeqnarray}{rCl} &&\sum_{\substack{w_1,w_2,\\v_1,v_2\\\text{with overlap}\quad \Theta=\left\{1,2\right\}}}{\Big( \Pr\big( F\left(w_1,w_2\right)\cap  
F\left(v_1,v_2\right)\big)\Big)}\nonumber\\&&=\sum_{w_1,w_2}{\Big( \Pr\big(F\left(w_1,w_2\right)\big)\Big)}\label{eq:case11}\nonumber\\&&<\sum_{w_1,w_2}\exp\Bigg(-n\bigg(I\Big(X;\hat{Y}^{\left(1\right)},\hat{Y}^{\left(2\right)},\hat{Y}^{\left(12\right)}\Big)\nonumber\\&& +I\Big(\hat{Y}^{\left(1\right)};\hat{Y}^{\left(2\right)}\Big)-4\epsilon_m\bigg)\Bigg)\label{eq:case12}\nonumber\\&&\leq\exp\Bigg(n\bigg(R_1+R_2+\epsilon_1+\epsilon_2\nonumber\\&&-I\Big(X;\hat{Y}^{\left(1\right)},\hat{Y}^{\left(2\right)},\hat{Y}^{\left(12\right)}\Big)-I\Big(\hat{Y}^{\left(1\right)};\hat{Y}^{\left(2\right)}\Big)\nonumber\end{IEEEeqnarray}\begin{IEEEeqnarray}{rCl}&&+\delta_2\bigg)\Bigg),\label{eq:case13}
\end{IEEEeqnarray}
where $\delta_2$ accounts for the $4\epsilon_m$-term and the offset from the flooring operations.
		
\textbf{Case $\mathbf{\Theta=\left\{1\right\}}$:} We have $w_1=v_1$ and $w_2\neq v_2$. 
Using the shorthands $\mathbb{A}_{\epsilon}^{\ast\left(n\right)}\big(\cdot|\mathbf x\big)$ for 
$\mathbb{A}_{\epsilon}^{\ast\left(n\right)}\big(p_{X,\hat{Y}^{\left(1\right)},\hat{Y}^{\left(2\right)},\hat{Y}^{\left(12\right)}}|\mathbf x\big) \quad \text{or} \quad \mathbb{A}_{\epsilon}^{\ast\left(n\right)}\big(p_{X,\hat{Y}^{\left(1\right)}}|\mathbf x\big)$, 
respectively (it should be clear from the context, which distribution needs to be plugged in), and $\mathbb{A}_{\epsilon}^{\ast\left(n\right)}\big(\cdot|\mathbf x,\hat{\mathbf y}^{\left(1\right)}\big)$ for	$\mathbb{A}_{\epsilon}^{\ast\left(n\right)}\big(p_{X,\hat{Y}^{\left(1\right)},\hat{Y}^{\left(2\right)},\hat{Y}^{\left(12\right)}}|\mathbf x,\hat{\mathbf y}^{\left(1\right)}\big),$
			we get 
\begin{IEEEeqnarray}{rCl}&&\Pr\big( F\left(w_1,w_2\right)\cap 
F\left(w_1,v_2\right)\big)\nonumber\\&=&\Pr \Bigg(\bigg\{\Big(\hat{\mathbf Y}^{\left(1\right)}\left(w_1\right),\hat{\mathbf Y}^{\left(2\right)}\left(w_2\right),\hat{\mathbf Y}^{\left(12\right)}\left(w_1,w_2\right)\Big)\nonumber\\&&\quad \quad \quad \quad \quad \quad \quad \quad \quad \quad \quad \quad  \in \mathbb{A}_{\epsilon}^{\ast\left(n\right)}\big(\cdot|\mathbf x\big) \bigg\}\nonumber\\&&\quad \quad\cap \bigg\{\Big(\hat{\mathbf Y}^{\left(1\right)}\left(w_1\right),\hat{\mathbf Y}^{\left(2\right)}\left(v_2\right),\hat{\mathbf Y}^{\left(12\right)}\left(w_1,v_2\right)\Big)\nonumber\\&&\quad \quad \quad \quad \quad \quad \quad \quad  \quad  \quad \quad \in \mathbb{A}_{\epsilon}^{\ast\left(n\right)}\big(\cdot|\mathbf x\big) \bigg\}\Bigg)\nonumber\label{eq:case21}\\&=&\Pr \Bigg(\bigg\{\hat{\mathbf Y}^{\left(1\right)}\left(w_1\right)\in\mathbb{A}_{\epsilon}^{\ast\left(n\right)}\big(\cdot|\mathbf x\big)\bigg\}\nonumber\\&&\quad \quad \cap \bigg\{\Big(\hat{\mathbf Y}^{\left(2\right)}\left(w_2\right),\hat{\mathbf Y}^{\left(12\right)}\left(w_1,w_2\right)\Big)\nonumber\\&& \quad \quad \quad \quad \quad \quad \quad \quad \in \mathbb{A}_{\epsilon}^{\ast\left(n\right)}\big(\cdot|\mathbf x,\hat{\mathbf Y}^{\left(1\right)}\left(w_1\right)\big)  \bigg\}\nonumber\\&&\quad \quad\cap \bigg\{\Big(\hat{\mathbf Y}^{\left(2\right)}\left(v_2\right),\hat{\mathbf Y}^{\left(12\right)}\left(w_1,v_2\right)\Big)\nonumber\\&& \quad \quad \quad \quad \quad \quad \in \mathbb{A}_{\epsilon}^{\ast\left(n\right)}\big(\cdot|\mathbf x,\hat{\mathbf Y}^{\left(1\right)}\left(w_1\right)\big)\bigg\} \Bigg)\label{eq:case22}\\&=& \Pr\left(\Lambda_a\cap\Lambda_b\cap\Lambda_c\right)\label{eq:case23}\\&=&\Pr\left(\Lambda_a\right)\Pr\left(\Lambda_b|\Lambda_a\right)\Pr\left(\Lambda_c|\Lambda_a,\Lambda_b\right),\label{eq:case24}	\end{IEEEeqnarray}
where in \eqref{eq:case22} we have used Lemma \ref{lem:chtypsets} (in Appendix); \eqref{eq:case23} must be understood as the definitions of the events $\Lambda_a$,$\Lambda_b$,$\Lambda_c$; and where \eqref{eq:case24} follows from the chain rule.
Note that conditionally on $\Lambda_a$, the events $\Lambda_b$ and $\Lambda_c$ are independent of each other, i.e., in \eqref{eq:case24} we have the two terms $\Pr\left(\Lambda_b|\Lambda_a\right)$ and $\Pr\left(\Lambda_c|\Lambda_a\right)$ that are basically the same. Let us investigate them more closely. We have
\begin{IEEEeqnarray}{rCl}&&\Pr\left(\Lambda_b|\Lambda_a\right)\nonumber\\&&=\sum_{\hat{y}^{\left(1\right)}\in \mathbb{A}_{\epsilon}^{\ast\left(n\right)}\left(\cdot|\mathbf x\right)}\Pr\left[\hat{\mathbf Y}^{\left(1\right)}\left(w_1\right)=\hat{\mathbf y}^{\left(1\right)}\big|\Lambda_a\right]\nonumber\\ 
		&&\cdot\Pr \Bigg[\Big(\hat{\mathbf Y}^{\left(2\right)}\left(w_2\right),\hat{\mathbf Y}^{\left(12\right)}\left(w_1,w_2\right)\Big)\in  \mathbb{A}_{\epsilon}^{\ast\left(n\right)}\left(\cdot|\mathbf x,\hat{\mathbf y}^{\left(1\right)}\right)\nonumber\\&&\quad \quad \quad \quad\Big|\hat{\mathbf Y}^{\left(1\right)}\left(w_1\right)=\hat{\mathbf y}^{\left(1\right)}\Bigg]\label{eq:case25}\nonumber\\&&\leq \underbrace{\sum_{\hat{y}^{\left(1\right)}\in \mathbb{A}_{\epsilon}^{\ast\left(n\right)}\left(\cdot|\mathbf x\right)}\Pr\left[\hat{\mathbf Y}^{\left(1\right)}\left(w_1\right)=\hat{\mathbf y}^{\left(1\right)}|\Lambda_a\right]}_{=1}\nonumber\end{IEEEeqnarray}\begin{IEEEeqnarray}{rCl}&&\cdot\max_{\substack{\hat{y}^{\left(1\right)}\in\\\ \mathbb{A}_{\epsilon}^{\ast\left(n\right)}\left(\cdot|\mathbf x\right)}}\Pr \Bigg[\Big(\hat{\mathbf Y}^{\left(2\right)}\left(w_2\right),\hat{\mathbf Y}^{\left(12\right)}\left(w_1,w_2\right)\Big)\nonumber\\&&\quad \quad \quad \quad \quad \quad \quad \in  \mathbb{A}_{\epsilon}^{\ast\left(n\right)}\left(\cdot|\mathbf x,\hat{\mathbf y}^{\left(1\right)}\right) \bigg\}\Bigg]\label{eq:case26}\\&&=\max_{\substack{\hat{y}^{\left(1\right)}\in\\  \mathbb{A}_{\epsilon}^{\ast\left(n\right)}\left(\cdot|\mathbf x\right)}}\Pr \Bigg[\Big(\hat{\mathbf Y}^{\left(2\right)}\left(w_2\right),\hat{\mathbf Y}^{\left(12\right)}\left(w_1,w_2\right)\Big)\nonumber\\&&\quad \quad \quad \quad \quad \quad \quad \in  \mathbb{A}_{\epsilon}^{\ast\left(n\right)}\left(\cdot|\mathbf x,\hat{\mathbf y}^{\left(1\right)}\right) \bigg\}\Bigg]\label{eq:case27}\\
&&=\max_{\substack{\hat{y}^{\left(1\right)}\in\\ \mathbb{A}_{\epsilon}^{\ast\left(n\right)}\left(\cdot|\mathbf x\right)}}\sum_{\substack{\left(\hat{\mathbf{y}}^{\left(2\right)},\hat{\mathbf{y}}^{\left(12\right)}\right)\in\\ \mathbb{A}_{\epsilon}^{\ast\left(n\right)}\left(\cdot|\mathbf x,\hat{\mathbf y}^{\left(1\right)}\right)}}p^n_{\hat{\mathbf Y}^{\left(2\right)}}\left(\hat{\mathbf y}^{\left(2\right)}\right)\nonumber\\&&\quad \quad \quad \quad \cdot p^n_{\hat{\mathbf Y}^{\left(12\right)}|\hat{\mathbf Y}^{\left(1\right)},\hat{\mathbf Y}^{\left(2\right)}}\left(\hat{\mathbf y}^{\left(12\right)}|\hat{\mathbf y}^{\left(1\right)},\hat{\mathbf y}^{\left(2\right)}\right)\label{eq:case28}\\
&&<\exp\Bigg(n\bigg(H\Big(\hat{Y}^{\left(2\right)},\hat{Y}^{\left(12\right)}|X,\hat{Y}^{\left(1\right)}\Big)-H\Big(\hat Y^{\left(2\right)}\Big)\nonumber\\&&\quad \quad \quad -H\Big(\hat{Y}^{\left(12\right)}|\hat{Y}^{\left(1\right)},\hat{Y}^{\left(2\right)}\Big)+3\epsilon_m\bigg)\Bigg).\label{eq:case211}
\end{IEEEeqnarray}
Here, \eqref{eq:case26}-\eqref{eq:case27} follows by replacing the average over $\hat{\mathbf y}^{\left(1\right)}$ by the maximum over $\hat{\mathbf y}^{\left(1\right)}$; in \eqref{eq:case28} we apply our knowledge about how the codebook has been generated; and the subsequent inequality follows again in the usual manner from Lemmas \ref{lem:TA}, \ref{lem:TB} (in Appendix). The same bound also applies to $\Pr\left(\Lambda_c|\Lambda_a,\Lambda_b\right)=\Pr\left(\Lambda_c|\Lambda_a\right)$. For $\Pr\left(\Lambda_a\right)$ we note that $\hat{\mathbf Y}^{\left(1\right)}\left(w_1\right)$ is generated completely independently of the source sequence $\mathbf X$. Hence we can apply Lemma \ref{lem:TC} (in Appendix):
\begin{equation}\Pr\left(\Lambda_a\right)<e^{-n\big(I\left(X;\hat Y^{\left(1\right)}\right)-2\epsilon_m\big)}
. \label{eq:case212}
\end{equation}
We plug \eqref{eq:case211} and \eqref{eq:case212} into \eqref{eq:case24} and get 
\begin{IEEEeqnarray}{rCl}&& \Pr\big( F\left(w_1,w_2\right)\cap 
	F\left(w_1,v_2\right)\big)\nonumber\label{eq:case213}
	\\&&<\exp\Bigg(-n\bigg(2I\big(X;\hat{ Y}^{\left(1\right)},\hat{Y}^{\left(2\right)},\hat{Y}^{\left(12\right)}\big)\nonumber\\&&\left.\left.+2I\big(\hat{Y}^{\left(1\right)};\hat{Y}^{\left(2\right)}\big)\right.\right.-I\big(X;\hat{Y}^{\left(1\right)}\big)-8\epsilon_m\bigg)\Bigg).\nonumber\label{eq:case218}
\end{IEEEeqnarray}
Hence, we get the following bound:
\begin{IEEEeqnarray}{rCl}
	&&\sum_{\substack{w_1,w_2,\\v_2\neq w_2}} \Pr\big( F\left(w_1,w_2\right)\cap  
	F\left(w_1,v_2\right)\big)\nonumber\\&&<\lfloor e^{n\left(R_1+\epsilon_1\right)}\rfloor\lfloor e^{n\left(R_2+\epsilon_2\right)}\rfloor\Big(\lfloor e^{n\left(R_2+\epsilon_1\right)}\rfloor-1\Big)
	\nonumber\\&&\cdot \exp\Bigg(-n\bigg(2I\big(X;\hat{ Y}^{\left(1\right)},\hat{Y}^{\left(2\right)},\hat{Y}^{\left(12\right)}\big)\nonumber\\&&\left.\left.+2I\big(\hat{Y}^{\left(1\right)};\hat{Y}^{\left(2\right)}\big)\right.\right.-I\big(X;\hat{Y}^{\left(1\right)}\big)-8\epsilon_m\bigg)\Bigg)\label{eq:case219}\nonumber\end{IEEEeqnarray}\begin{IEEEeqnarray}{rCl}&&\leq \exp\Bigg(n\bigg(R_1+2R_2+\epsilon_1+2\epsilon_2\nonumber\\&&-2I\big(X;\hat{ Y}^{\left(1\right)},\hat{Y}^{\left(2\right)},\hat{Y}^{\left(12\right)}\big)\nonumber\\&&\left.\left.-2I\big(\hat{Y}^{\left(1\right)};\hat{Y}^{\left(2\right)}\big)\right.\right.+I\big(X;\hat{Y}^{\left(1\right)}\big)+\delta_3\bigg)\Bigg),\label{eq:case220}
\end{IEEEeqnarray}
where $\delta_3$ accounts for $8\epsilon_m$ and the rounding offsets.

\textbf{Case $\mathbf{\Theta=\left\{2\right\}}$:} This is the case as the case $\Theta=\left\{1\right\}$, but with exchanged roles of $\hat{Y}^{\left(1\right)}$ and  $\hat{Y}^{\left(2\right)}$:
\begin{IEEEeqnarray}{rCl}&&\sum_{\substack{w_1,w_2,\\v_1\neq w_1}} \Pr\big( F\left(w_1,w_2\right)\cap 
	F\left(v_1,w_2\right)\big)\nonumber\\&<& \exp\Bigg(n\bigg(2R_1+R_2+2\epsilon_1+\epsilon_2\nonumber\\&&-2I\big(X;\hat{ Y}^{\left(1\right)},\hat{Y}^{\left(2\right)},\hat{Y}^{\left(12\right)}\big)\nonumber\\&&\left.\left.-2I\big(\hat{Y}^{\left(1\right)};\hat{Y}^{\left(2\right)}\big)\right.\right.+I\big(X;\hat{Y}^{\left(2\right)}\big)+\delta_4\bigg)\Bigg).\label{eq:case31}
\end{IEEEeqnarray}
\textbf{Case $\mathbf{\Theta=\emptyset}$:} In this case we have both $w_1\neq v_1$ and $w_2\neq v_2$, i.e., the two events $
F\left(w_1,w_2\right)$ and  $
F\left(v_1,v_2\right)$ are independent. Hence, we have
\begin{IEEEeqnarray}{rCl}&&\sum_{\substack{w_1,w_2,v_1,v_2\nonumber\\\text{with no overlap}}} \Pr\big( F\left(w_1,w_2\right)\cap 
	F\left(v_1,v_2\right)\big)\label{eq:case41}\nonumber\\&&=\sum_{w_1,w_2}\sum_{\substack{v_1\neq w_1\\v_2\neq w_1}} \Pr\big( F\left(w_1,w_2\right)\big)\Pr\big( F\left(v_1,v_2\right)\big)\label{eq:case42}\nonumber\\&&=\sum_{w_1,w_2}\Pr\big( F\left(w_1,w_2\right)\big)\sum_{\substack{v_1\neq w_1\\v_2\neq w_1}}\Pr\big( F\left(v_1,v_2\right)\big)\label{eq:case43}\nonumber\\&&\leq \sum_{w_1,w_2}\Pr\big( F\left(w_1,w_2\right)\big)\sum_{v_1,v_2}\Pr\big( F\left(v_1,v_2\right)\big)\label{eq:case44}\\
	&&=\Big(\sum_{w_1,w_2}\Pr\big( F\left(w_1,w_2\right)\big)\Big)^2=\left(\mathbb E\left[K\right]\right)^2. \label{eq:case45}
	\end{IEEEeqnarray}
Here, in \eqref{eq:case44} we increase the number of terms in the second sum; and in \eqref{eq:case45} we use  \eqref{eq:kapamean3}. Hence, by plugging  \eqref{eq:case13},\eqref{eq:case220},\eqref{eq:case31},\eqref{eq:case45} in \eqref{eq:skapamean3}, and using \eqref{eq:kapamean9}, we get from \eqref{eq:tprobability} the following bound for $\Pr\left(\text{Case b}\right)$: 
\begin{IEEEeqnarray*}{rCl}
	&&\Pr\left(\text{Case b}\right)
	<4\exp\Bigg(n\bigg(-R_1-R_2-\epsilon_1-\epsilon_2\nonumber\\&&+I\Big(X;\hat{Y}^{\left(1\right)},\hat{Y}^{\left(2\right)},\hat{Y}^{\left(12\right)}\Big) +I\Big(\hat{Y}^{\left(1\right)};\hat{Y}^{\left(2\right)}\Big)\\&&+\delta_2+2\delta_1\bigg)\Bigg)+4\exp\Bigg(n\bigg(-R_1+\epsilon_1
	+I\big(X;\hat{Y}^{\left(1\right)}\big)\\&&+\delta_3+2\delta_1\bigg)\Bigg) +4\exp\Bigg(n\bigg(-R_2-\epsilon_2\nonumber+I\big(X;\hat{Y}^{\left(2\right)}\big)\\&&+\delta_4+2\delta_1\bigg)\Bigg)\triangleq \delta_5,
\end{IEEEeqnarray*}
where $\delta_5$ is arbitrarily small if $n$ is large enough and $\epsilon$ small enough such that
$\delta_2+2\delta_1<\epsilon_1+\epsilon_2,$\newline$\delta_3+2\delta_1<\epsilon_1, \delta_4+2\delta_1<\epsilon_2,$
and if 
$R_1\geq I\left(X;\hat{Y}^{\left(1\right)}\right)$,
$R_2\geq I\left(X;\hat{Y}^{\left(2\right)}\right)$,
$R_1+R_2\geq I\left(X;\hat{Y}^{\left(1\right)},\hat{Y}^{\left(2\right)},\hat{Y}^{\left(12\right)}\right)+I\left(\hat{Y}^{\left(1\right)};\hat{Y}^{\left(2\right)}\right)$.
Putting all the cases back into \eqref{eq:totalE}, now gives us the characterization of the Theorem. This completes the proof.
\end{itemize}	

\subsection{Proof of Theorem \ref{th:theorem2}}

Assume that there exists a third encoder whose index always safely arrives at the decoder. i.e., the third encoder sees a noise-free channel. We assign the rate $R_0$ to this third encoder. 
\begin{itemize}
	\item{\textit{Setup}:} Choose $\epsilon_0,\epsilon_1,\epsilon_2>0$, a PMF $p_{\hat{Y}^{\left(0\right)},\hat{Y}^{\left(1\right)},\hat{Y}^{\left(2\right)},\hat{Y}^{\left(12\right)}|X}$, compute the marginals $p_{\hat{Y}^{\left(1\right)}|\hat{Y}^{\left(0\right)}}$,  $p_{\hat{Y}^{\left(2\right)}|\hat{Y}^{\left(0\right)}}$ and $p_{\hat{Y}^{\left(12\right)}|\hat{Y}^{\left(0\right)},\hat{Y}^{\left(1\right)},\hat{Y}^{\left(2\right)}}$ as marginal distributions of $p_Xp_{\hat{Y}^{\left(0\right)},\hat{Y}^{\left(1\right)},\hat{Y}^{\left(2\right)},\hat{Y}^{\left(12\right)}|X}$.
	\item{\textit{Codebook design}:} We independently generate $\lfloor e^{n\left(R_0+\epsilon_0\right)} \rfloor$ length-$n$ codewords 
	$\hat{\mathbf Y}^{\left(0\right)}\left(w_0\right), w_0=1,\dots, \lfloor e^{n\left(R_0+\epsilon_0\right)} \rfloor$, by choosing each of the $n\lfloor e^{n\left(R_0+\epsilon_0\right)} \rfloor$ symbols $\hat{Y}^{\left(0\right)}_k\left(w_0\right)$ independently at random according to $p_{\hat{Y}^{\left(0\right)}}$. For every $w_0$, we independently generate  $\lfloor e^{n\left(R_i+\epsilon_i\right)} \rfloor$ length-n codewords 
	\begin{multline*}
	\hat{\mathbf Y}^{\left(i\right)}\left(w_0,w_i\right)\sim p^n_{\hat{Y}^{\left(i\right)}|\hat{ Y}^{\left(0\right)}}\big(\cdot|\hat{\mathbf Y}^{\left(0\right)}\left(w_i\right)\big),  \\w_i=1,\dots, \lfloor e^{n\left(R_i+\epsilon_i\right)} \rfloor,
	\end{multline*}
	for $i=1,2$
	(this means that we have $\lfloor e^{n\left(R_0+\epsilon_0\right)} \rfloor\cdot \lfloor e^{n\left(R_1+\epsilon_1\right)} \rfloor$ codewords $\hat{\mathbf Y}^{\left(1\right)}$ and $\lfloor e^{n\left(R_0+\epsilon_0\right)} \rfloor\cdot \lfloor e^{n\left(R_2+\epsilon_2\right)} \rfloor$ codewords $\hat{\mathbf Y}^{\left(2\right)}$). 
	Finally, for each triple $\left(w_0,w_1,w_2\right)$, we generate one length-$n$ codeword
	\begin{IEEEeqnarray*}{rCl} &&\hat{\mathbf Y}^{\left(12\right)}\left(w_0,w_1,w_2\right)\sim p^n_{\hat{ Y}^{\left(12\right)}|\hat{ Y}^{\left(0\right)},\hat{ Y}^{\left(1\right)},\hat{ Y}^{\left(2\right)}}\big(\cdot|\hat{\mathbf Y}^{\left(0\right)}\left(w_0\right)\\&&\quad \quad \quad \quad \quad \quad \quad \quad,\hat{\mathbf Y}^{\left(1\right)}\left(w_0,w_1\right),\hat{\mathbf Y}^{\left(0\right)}\left(w_0,w_2\right)\big).
	\end{IEEEeqnarray*}
	\item{\textit{Encoder Design}:} For a given sequence $\mathbf{x}$, the encoders try to find a triple $\left(w_0,w_1,w_2\right)$ such that 
	\begin{IEEEeqnarray}{rCl}
		&&\left(\mathbf x,\hat{\mathbf Y}^{\left(0\right)}\left(w_0\right),\hat{\mathbf Y}^{\left(1\right)}\left(w_0,w_1\right),\hat{\mathbf Y}^{\left(2\right)}\left(w_0,w_2\right)\right.\nonumber\\&& \left.,\hat{\mathbf Y}^{\left(12\right)}\left(w_0,w_1,w_2\right)\right)\nonumber\\&&\quad \quad \quad \in \mathbb{A}_{\epsilon}^{\ast\left(n\right)}\big(p_{X,\hat{Y}^{\left(0\right)},\hat{Y}^{\left(1\right)},\hat{Y}^{\left(2\right)},\hat{Y}^{\left(12\right)}}\big).\IEEEeqnarraynumspace \label{jtypicality2}
	\end{IEEEeqnarray}
	\item{\textit{Decoder Design}:} The decoder still consists of only three different decoding functions $\left(y_1^n,y_2^n,y_{12}^n\right)$, because $w_0$ arrives for sure and we are not interested in the case when only $w_0$ arrives (this still counts like nothing has arrived identically to the set-up described before). The decoder puts out $\hat{\mathbf Y}^{\left(i\right)}\left(w_0,w_i\right)$ if  $\left(w_0, w_i\right)$ is received for $i=1,2,12$.
	\item{\textit{Performance Analysis}:} We again distinguish three different cases: (a) $\mathbf x\notin \mathbb{A}_{\epsilon^\prime}^{\ast\left(n\right)}\big(p_{X}\big)$ (b) $\mathbf x \in \mathbb{A}_{\epsilon^\prime}^{\ast\left(n\right)}\big(p_{X}\big)$ and for every pair $\left(w_0,w_1,w_2\right)$, \eqref{jtypicality2} is not satisfied (c) $\mathbf x\in \mathbb{A}_{\epsilon^\prime}^{\ast\left(n\right)}\big(p_{X}\big)$ and exists a pair $\left(w_0,w_1,w_2\right)$ such that \eqref{jtypicality2} is satisfied. The analysis of (a) and (c) is similar to what we did in Theorem \ref{th:theorem1}, thus, we omit it. We only study $\Pr\left(\text{Case b}\right)$.
		Define
		\begin{IEEEeqnarray*}{rCl}&&F\left(w_0,w_1,w_2\right)\triangleq\\&& \Bigg\{\left(\hat{\mathbf Y}^{\left(0\right)}\left(w_0\right),\hat{\mathbf Y}^{\left(1\right)}\left(w_0,w_1\right),\hat{\mathbf Y}^{\left(2\right)}\left(w_0,w_2\right),\right.\\&& \left. \hat{\mathbf Y}^{\left(12\right)}\left(w_0,w_1,w_2\right)\right)\\&&\in \mathbb{A}_{\epsilon}^{\ast\left(n\right)}\big(p_{X,\hat{Y}^{\left(0\right)},\hat{Y}^{\left(1\right)},\hat{Y}^{\left(2\right)},\hat{Y}^{\left(12\right)}}|\mathbf x\big)\Bigg\},\end{IEEEeqnarray*}
		and we use the same trick on the sum of the indicator functions $K$:
	\begin{IEEEeqnarray}{rCl}
	&&\mathbb E[K]
	>\exp\Bigg(n\bigg(R_0+R_1+R_2+\epsilon_0+\epsilon_1+\epsilon_2 \nonumber\\&& \quad-I\Big(X;\hat{Y}^{\left(0\right)},\hat{Y}^{\left(1\right)},\hat{Y}^{\left(2\right)},\hat{Y}^{\left(12\right)}\Big)\nonumber\\&& \quad-I\Big(\hat{Y}^{\left(1\right)};\hat{Y}^{\left(2\right)}|\hat{Y}^{\left(0\right)}\Big)-\delta_1\bigg)\Bigg),\IEEEeqnarraynumspace \IEEEeqnarraynumspace\label{eq:kapameane}
		\end{IEEEeqnarray}
		\begin{IEEEeqnarray}{rCl}&&\mathbb  E\left[K^2\right]\nonumber\\&&=\sum_{\substack{\Theta\\\subseteq\left\{0,1,2\right\}}}\sum_{\substack{w_0,w_1,w_2,\\v_0,v_1,v_2\\ \text{with}\\\text{overlap}\\\Theta}}\Pr\Big( F\left(w_0,w_1,w_2\right)\nonumber\\&&\quad \quad \quad \quad \quad \quad \quad \quad \quad \quad   \cap F\left(v_0,v_1,v_2\right)\Big).\IEEEeqnarraynumspace\label{eq:kapameane1}
		\end{IEEEeqnarray}
	     If $w_0\neq v_0$, then, the two events $F\left(w_0,w_1,w_2\right)$ and $F\left(v_0,v_1,v_2\right)$ are disjoint because $w_0$ is a counter that was used in the generation of all codewords simultaneously so the cases $\Theta=\left\{1,2\right\},\Theta=\left\{1\right\}, \Theta=\left\{2\right\}$ and $\Theta=\emptyset$ can be treated jointly.
	 	 
	 	 \textbf{Case $\mathbf{\Theta=\left\{1,2\right\},\Theta=\left\{1\right\},\Theta=\left\{2\right\}}$ and $\mathbf{\Theta=\emptyset}$:} 
		\begin{IEEEeqnarray}{rCl} &&\sum_{\substack{w_0,w_1,w_2,\\v_0,v_1,v_2\\ \text{with no overlap}\\\text{in $w_0$}}} \Pr\big(F\left(w_0,w_1,w_2\right)\cap 
			F\left(v_0,v_1,v_2\right)\big)\nonumber\\&&= \sum_{\substack{w_0,w_1,\\w_2}} \Pr\big(F\left(w_0,w_1,w_2\right)\big)\nonumber\\&& \cdot \sum_{\substack{v_0\neq w_0,\\v_1,v_2}} \Pr\big( F\left(v_0,v_1,v_2\right)  
			\big)\nonumber\\&&\leq \bigg(\sum_{w_0,w_1,w_2} \Pr\big(F\left(w_0,w_1,w_2\right)\big)\bigg)^2\nonumber\\&&=\left(\E\left[K\right]\right)^2.\label{eq:case1e}
		\end{IEEEeqnarray}
		\textbf{Case $\mathbf{\Theta=\left\{0,1,2\right\}}$:}
		\begin{IEEEeqnarray}{rCl}&&\sum_{\substack{w_0,w_1,w_2,\\v_0,v_1,v_2\\ \text{with overlap}\\\text{$\left\{0,1,2\right\}$}}} \Pr\big(F\left(w_0,w_1,w_2\right)\cap  
			F\left(v_1,v_1,v_2\right)\big)\nonumber\\&&= \sum_{w_0,w_1,w_2} \Pr\big(F\left(w_0,w_1,w_2\right)\big)\nonumber\end{IEEEeqnarray}\begin{IEEEeqnarray}{rCl}&&<\exp\Bigg(n\bigg(R_0+R_1+R_2+\epsilon_0+\epsilon_1+\epsilon_2\nonumber\\&&\quad \quad  \quad \quad-I\Big(X;\hat{Y}^{\left(0\right)},\hat{Y}^{\left(1\right)},\hat{Y}^{\left(2\right)},\hat{Y}^{\left(12\right)}\Big)\nonumber\\&&\quad \quad \quad \quad \quad-I\Big(\hat{Y}^{\left(1\right)};\hat{Y}^{\left(2\right)}|\hat{Y}^{\left(0\right)}\Big)+\delta_2\bigg)\Bigg).\IEEEeqnarraynumspace\label{eq:case2e}
		\end{IEEEeqnarray}
		\textbf{Case $\mathbf{\Theta=\left\{0,1\right\}}$:}
		\begin{IEEEeqnarray}{rCl} &&\sum_{\substack{w_0,w_1,w_2,v_0,v_1,v_2\\ \text{with overlap $\left\{0,1\right\}$}}} \Pr\big(F\left(w_0,w_1,w_2\right)\cap 
			F\left(v_1,v_1,v_2\right)\big)\nonumber\\
			&\leq& \exp\Bigg(n\bigg(R_0+R_1+2R_2+\epsilon_0+\epsilon_1+2\epsilon_2\nonumber\\&&+I\big(X;\hat{ Y}^{\left(0\right)},\hat{Y}^{\left(1\right)}\big)-2I\big(X;\hat{ Y}^{\left(0\right)},\hat{ Y}^{\left(1\right)},\hat{Y}^{\left(2\right)},\hat{Y}^{\left(12\right)}\big)\nonumber\\&&-2I\big(\hat{Y}^{\left(1\right)};\hat{Y}^{\left(2\right)}|\hat{Y}^{\left(0\right)}\big)+\delta_3\bigg)\Bigg). \label{eq:case3e}
		\end{IEEEeqnarray}
		\textbf{Case $\mathbf{\Theta=\left\{0,2\right\}}$:} 
		\begin{IEEEeqnarray}{rCl}&&
			\sum_{\substack{w_0,w_1,w_2,v_0,v_1,v_2\\ \text{with overlap $\left\{0,2\right\}$}}} \Pr\big(F\left(w_0,w_1,w_2\right)\cap 
			F\left(v_1,v_1,v_2\right)\big)\nonumber\\&&<\exp\Bigg(n\bigg(R_0+2R_1+R_2+\epsilon_0+2\epsilon_1+\epsilon_2\nonumber\\&&+I\big(X;\hat{ Y}^{\left(0\right)},\hat{ Y}^{\left(2\right)}\big)-2I\big(X;\hat{ Y}^{\left(0\right)},\hat{ Y}^{\left(1\right)},\hat{Y}^{\left(2\right)},\hat{Y}^{\left(12\right)}\big)\nonumber\\&&-2I\big(\hat{Y}^{\left(1\right)};\hat{Y}^{\left(2\right)}|\hat{Y}^{\left(0\right)}\big)+\delta_4\bigg)\Bigg).\label{eq:case4e}
		\end{IEEEeqnarray}
		\textbf{Case $\mathbf{\Theta=\left\{0\right\}}$:} 
		\begin{IEEEeqnarray}{rCl}&&\sum_{\substack{w_0,w_1,w_2,v_0,v_1,v_2 \nonumber\\ \text{with overlap $\left\{0\right\}$}}} \Pr\big( F\left(w_0,w_1,w_2\right)\cap  
			F\left(v_0,v_1,v_2\right)\big)\nonumber\\
			&&< \exp\Bigg(n\bigg(R_0+2R_1+2R_2+I\big(X;\hat{ Y}^{\left(0\right)}\big)\nonumber\\&&-2I\big(X;\hat{ Y}^{\left(0\right)},\hat{ Y}^{\left(1\right)},\hat{Y}^{\left(2\right)},\hat{Y}^{\left(12\right)}\big)\nonumber\\&&-2I\big(\hat{Y}^{\left(1\right)};\hat{Y}^{\left(2\right)}|\hat{Y}^{\left(0\right)}\big)+\delta_5\bigg)\Bigg). \label{eq:case5e}
			\end{IEEEeqnarray}
		   Next, we plug \eqref{eq:case1e},\eqref{eq:case2e},\eqref{eq:case3e},\eqref{eq:case4e},\eqref{eq:case5e} in \eqref{eq:kapameane1} and using \eqref{eq:kapameane} we obtain: 
			\begin{IEEEeqnarray}{rCl}
			&&\Pr\left(\text{Case b}\right)\leq\frac{\Big(\mathbb E\left[K^2\right]-\left(\mathbb E\left[K\right]\right)^2\Big)}{\left(\mathbb E\left[K\right]\right)^2}\nonumber\\
			&&<4\exp\Bigg(n\bigg(-R_0-R_1-R_2-\epsilon_0-\epsilon_1-2\epsilon_2\nonumber\\&&+\delta_2+2\delta_1+I\Big(X;\hat{Y}^{\left(0\right)},\hat{Y}^{\left(1\right)},\hat{Y}^{\left(2\right)},\hat{Y}^{\left(12\right)}\Big)\nonumber\\&& +I\Big(\hat{Y}^{\left(1\right)};\hat{Y}^{\left(2\right)}|\hat{Y}^{\left(0\right)}\Big)\bigg)\Bigg)+4\exp\Bigg(n\bigg(-R_0\nonumber\\
				&&-R_1-\epsilon_0-\epsilon_1+I\big(X;\hat{Y}^{\left(0\right)},\hat{Y}^{\left(1\right)}\big)+\delta_3+2\delta_1\bigg)\Bigg)\nonumber\end{IEEEeqnarray}\begin{IEEEeqnarray}{rCl}&& +4\exp\Bigg(n\bigg(-R_0-R_2-\epsilon_0-\epsilon_2\nonumber\nonumber\\&&+I\big(X;\hat{Y}^{\left(0\right)},\hat{Y}^{\left(2\right)}\big)+\delta_4+2\delta_1\bigg)\Bigg)\nonumber\\&& +4\exp\Bigg(n\bigg(-R_0-\epsilon_0+I\left(X;Y^{\left(0\right)}\right)\nonumber\\&&+\delta_5+2\delta_1\bigg)\Bigg) \triangleq \delta_6,
			\end{IEEEeqnarray}
			where $\delta_6$ is arbitrarily small if $n$ is large enough and $\epsilon$ small enough such that
			$\delta_2+2\delta_1<\epsilon_0+\epsilon_1+\epsilon_2$, $\delta_3+2\delta_1<\epsilon_0+\epsilon_1$, $ \delta_4+2\delta_1<\epsilon_0+\epsilon_2$,                      $\delta_5+2\delta_1<\epsilon_0$,
			and if $R_0\geq I\left(X,\hat{Y}^{\left(0\right)}\right)$, $
			R_0+R_1>I\left(X;\hat{Y}^{\left(0\right)},\hat{Y}^{\left(1\right)}\right)$, $
			R_0+R_2\geq I\left(X;\hat{Y}^{\left(0\right)},\hat{Y}^{\left(2\right)}\right)$, $ 
			 R_0+R_1+R_2\geq I\left(X;\hat{Y}^{\left(0\right)},\hat{Y}^{\left(1\right)},\hat{Y}^{\left(2\right)},\hat{Y}^{\left(12\right)}\right)+I\left(\hat{Y}^{\left(1\right)};\hat{Y}^{\left(2\right)}|\hat{Y}^{\left(0\right)}\right)$. In fact, we do not have access to such a third guaranteed channel. However, we can simulate it by adding $nR_0$ nats to both of the two other channels. Then, in all interesting three cases (only $w_1$ arrives, only $w_2$ arrives, and both arrive) we have these nats and they act like they had come over the virtual third channel. This now means that we adapt our rates $R_1$ and $R_2$:
			$\tilde{R}_1\triangleq R_1+R_0$ and $\tilde{R}_2\triangleq R_2+R_0$. Plugging this into our relationships, using that $R_0\geq I\left(X;\hat{Y}^{\left(0\right)}\right)$ and rename $\hat{Y}^{\left(0\right)}$ as $U$, in combination will the other two cases, give us the characterization of the Theorem. This completes the proof.
			\end{itemize}	
			
			\section{Appendix}
		    \begin{definition}[Strongly $\epsilon$-typical set \cite{moser:2019}]
			Fix an $\epsilon>0$, a PMF $p_{X,Y}\left(x,y\right)$, and a blocklength n. The strongly typical set $\mathbb{A}_{\epsilon}^{\ast\left(n\right)}\left(p_{X,Y}\right)$ with respect to the PMF $p_{X,Y}\left(x,y\right)$ is defined as 
			\begin{multline*}
			\mathbb{A}_{\epsilon}^{\ast\left(n\right)}\left(p_{X,Y}\right)\triangleq\\
			\left.
			\left\{ \,
			\begin{IEEEeqnarraybox}[
			\IEEEeqnarraystrutmode
			\IEEEeqnarraystrutsizeadd{2pt}
			{2pt}][c]{l}
			\left(\mathbf x,\mathbf y\right)\in \mathbb X^n\times \mathbb Y^n:\\
			|P_{\mathbf x,\mathbf y}\left(a,b\right)-p_{X,Y}\left(a,b\right)|<\frac{\epsilon}{|\mathbb X||\mathbb Y|},\forall \left(a,b\right)\in \mathbb X\times \mathbb Y, \text{and}\\ P_{\mathbf x,\mathbf y}\left(a,b\right)=0 ,\forall \left(a,b\right)\in \mathbb X\times \mathbb Y \quad \text{with} \quad p_{X,Y}\left(a,b\right)=0
			\end{IEEEeqnarraybox}\right\}.
			\right.
			\label{def:typsets}
			\end{multline*}
			
		\end{definition}
		\begin{definition}[Conditionally $\epsilon$-typical set  \cite{moser:2019}]
			For some joint PMF $p_{X,Y}$ with marginal $p_X$ and for some given sequence $\mathbf x \in \mathbb{A}_{\epsilon}^{\ast\left(n\right)}\left(p_X\right)$, we define the conditionally strongly typical set with respect to $p_{X,Y}$ as 
			\begin{IEEEeqnarray*}{rCl}&&\mathbb{A}_{\epsilon}^{\ast\left(n\right)}\big(p_{X,Y}|\mathbf x\big)\triangleq \left\{\mathbf y\in \mathbb Y^n:\left(\mathbf x,\mathbf y\right)\in \mathbb{A}_{\epsilon}^{\ast\left(n\right)}\big(p_{X,Y}\big) \right\}.
			\end{IEEEeqnarray*}
			\label{def:ctypsets}
		\end{definition}
	\begin{lemma}[\cite{moser:2019}]The event $\Big\{\left(\mathbf X,\mathbf Y\right)\in \mathbb{A}_{\epsilon}^{\ast\left(n\right)}\left(p_{X,Y}\right) \Big\}$
		is equivalent to the event 
		\begin{IEEEeqnarray*}{rCl}&&
		\left\{\mathbf X\in \mathbb{A}_{\epsilon}^{\ast\left(n\right)}\left(p_{X}\right) \right\}\cap \left\{\mathbf Y\in \mathbb{A}_{\epsilon}^{\ast\left(n\right)}\big(p_{X,Y}|\mathbf X\big)\right\}.
		\end{IEEEeqnarray*}\label{lem:chtypsets}
	\end{lemma}
\begin{definition}[$\epsilon_m,\delta_t$  \cite{moser:2019}]
	We name one particular $\epsilon$ and one particular $\delta$ that we meet often:
	\begin{IEEEeqnarray*}{rCl}&&\epsilon_m\big(p_{X,Y}\left(x,y\right)\big)\triangleq-\epsilon\log\left(p_{X,Y}^{\min}\right),\\&&\delta_{t}\left(n,\epsilon,\mathbb{X}\times \mathbb{Y} \right)\triangleq \left(n+1\right)^{|\mathbb X||\mathbb Y|}e^{-n\frac{\epsilon^2}{2|\mathbb X|^2|\mathbb Y|^2}\log e},
	\end{IEEEeqnarray*}
	where $p_{X,Y}^{\min}$ denotes the smallest positive value of $p_{X,Y}\left(x,y\right)$.
	\label{def:epsilondelta}
\end{definition}
\begin{lemma}[\cite{moser:2019}] Let $\left(\mathbf x,\mathbf y\right) \in \mathbb{A}_{\epsilon}^{\ast\left(n\right)}\left(p_{X,Y}\right)$. Then,
	\begin{IEEEeqnarray*}{rCl}&&e^{-n\Big(H\left(X,Y\right)+\epsilon_m\big(p_{X,Y}\left(x,y\right)\big)\Big)}< p_{X,Y}^n\left(\mathbf x,\mathbf y\right)\\&&< e^{-n\Big(H\left(X,Y\right)-\epsilon_m\big(p_{X,Y}\left(x,y\right)\big)\Big)}.\end{IEEEeqnarray*}
	Moreover, \begin{IEEEeqnarray*}{rCl}&&1-\delta_{t}\left(n,\epsilon,\mathbb{X}\times \mathbb{Y}\right)\leq\Pr\left[\left(\mathbf x, \mathbf y\right) \in \mathbb{A}_{\epsilon}^{\ast\left(n\right)}\left(p_{X,Y}\right)\right]\leq 1.\end{IEEEeqnarray*}
	\label{lem:TA}
	\end{lemma}	
	\begin{lemma}[\cite{moser:2019}]
 For every $\mathbf y\in \mathbb{A}_{\epsilon}^{\ast\left(n\right)}\left(p_{X,Y}|\mathbf x\right)$ we obtain 
	\begin{IEEEeqnarray*}{rCl}&&
	e^{-n\Big(H\left(Y|X\right)+\epsilon_m\big(p_{X,Y}\left(x,y\right)\big)\Big)}< p_{Y|X}^n\left(\mathbf y|\mathbf x\right)\\&&< e^{-n\Big(H\left(Y|X\right)-\epsilon_m\big(p_{X,Y}\left(x,y\right)\big)\Big)}.
	\end{IEEEeqnarray*}
	The size of the conditionally strongly typical set is bounded as 
	\begin{IEEEeqnarray*}{rCl}&&|\mathbb{A}_{\epsilon}^{\ast\left(n\right)}\big(p_{X,Y}|\mathbf x\big)|< e^{n\Big(H\left(Y|X\right)+\epsilon_m\big(p_{X,Y}\left(x,y\right)\big)\Big)}.\end{IEEEeqnarray*}
	If also $\mathbf x \in\mathbb{A}_{\frac{\epsilon}{2|\mathbb Y|}}^{\ast\left(n\right)}\left(p_{X}\right)$, then, we can obtain 
	\begin{IEEEeqnarray*}{rCl}&&|\mathbb{A}_{\epsilon}^{\ast\left(n\right)}\big(p_{X,Y}|\mathbf x\big)|> \left(1-\delta_{t}\big(n,\frac{\epsilon}{2},\mathbb{X}\times \mathbb{Y} \big)\right)\\&&\cdot e^{n\Big(H\left(Y|X\right)-\epsilon_m\big(p_{X,Y}\left(x,y\right)\big)\Big)}.\end{IEEEeqnarray*}

	\label{lem:TB}
	\end{lemma}
	\begin{lemma}[\cite{moser:2019}]
	Let $p_{X,Y}\left(x,y\right)$ be a joint PMF with marginals $p_X\left(x\right), p_Y\left(y\right)$. Let $\left(\mathbf x,\mathbf y\right)$ be generated:$\left\{\left(x_k,y_k\right)\right\}_{k=1}^n \text{IID} \sim p_X\left(x\right)p_Y\left(y\right).$
	Then,
	\begin{IEEEeqnarray*}{rCl}&&\big(1-\delta_{t}\left(n,\epsilon,\mathbb{X}\times \mathbb{Y} \right)\big)e^{-n\big(I\left(X;Y\right)+\epsilon_2\big)}\\&&<\Pr\left[\left(\mathbf x,\mathbf y\right) \in \mathbb{A}_{\epsilon}^{\ast\left(n\right)}\left(p_{X,Y}\right)\right]<e^{-n\big(I\left(X;Y\right)-\epsilon_2\big)},\end{IEEEeqnarray*} where 
	$\epsilon_2\triangleq \epsilon_m\big(p_{X,Y}\left(x,y\right)\big)+\epsilon_m\big(p_X\left(x\right)\big)+\epsilon_m\big(p_Y\left(y\right)\big)\leq 3\epsilon_m\big(p_{X,Y}\left(x,y\right)\big).$
	
	Moreover,
	\begin{IEEEeqnarray*}{rCl}&&
	\Pr\left[\mathbf Y\in \mathbb{A}_{\epsilon}^{\ast\left(n\right)}\big(p_{X,Y}|\mathbf x\big)\right]<e^{-n\big(I\left(X;Y\right)-\epsilon_3\big)},
	\end{IEEEeqnarray*}
	where $\epsilon_3\triangleq \epsilon_m\big(p_{X,Y}\left(x,y\right)\big)+\epsilon_m\big(p_Y\left(y\right)\big)\leq 2\epsilon_m\big(p_{X,Y}\left(x,y\right)\big)$.
    Further, if $\mathbf x \in \mathbb{A}_{\frac{\epsilon}{2|\mathbb Y|}}^{\ast\left(n\right)}\left(p_{X}\right)$, then,  we obtain 
	\begin{IEEEeqnarray*}{rCl}
	&&\Pr\left[\mathbf Y\in \mathbb{A}_{\epsilon}^{\ast\left(n\right)}\big(p_{X,Y}|\mathbf x\big)\right]\\&&>\big(1-\delta_t\left(n,\frac{\epsilon}{2},\mathbb X\times \mathbb Y\right)\big) e^{-n\big(I\left(X;Y\right)+\epsilon_3\big)}.
	\end{IEEEeqnarray*}
	
\label{lem:TC}
	\end{lemma}
\bibliographystyle{IEEEtran}
\bibliography{string,references}
\end{document}